**Realistic quantum device data synthesized by consumer AI and how to identify it**

S. M. Frolov [1] and O. V. Kravchenko [2,3]

[1] Department of Physics and Astronomy, University of Pittsburgh, Pittsburgh PA, USA
[2] Department of Family Medicine, University of Pittsburgh, Pittsburgh PA, USA
[3] Department of Biomedical Informatics, University of Pittsburgh, Pittsburgh PA, USA

**Abstract**

With the advance of generative artificial intelligence (AI) synthetic texts and images have become commonplace. These capabilities offer clear benefits, but have also raised a number of ethical concerns that often have to do with misrepresenting AI outputs as genuine material. A lesser known capability of generative AI is to perform the basic analysis, processing and even synthesis of numerical data. This raises the question of whether AI can be used to imitate experimental data that an expert would consider scientifically meaningful and on par with data in the figures of peer-reviewed manuscripts? In this paper, we focus on synthesizing data inspired by well-known experiments done frequently on quantum electronic devices. This field is related to information technologies such as spintronics and quantum computing, and is considered data-rich and data-driven. We demonstrate that it is possible to generate dramatic signals associated with iconic effects such as quantum bit control, Majorana fermions, Josephson effects, quantum dots and wires using widely available ChatGPT. We find that because some of the clearest data from quantum devices can be expressed in terms of relatively basic mathematical models, AI does not need to learn on the specialized body of data. Instead, knowledge of the physics equations and of the basic features of experimental signals can go a long way towards building a realistic dataset. We also demonstrate that real data can be augmented by AI, and that AI can mimic the noise of common scientific instruments. To help assure that published data come from experiments and are not synthesized by AI, we recommend sharing large volumes of the primary data. While it is straightforward for AI to mimic a few sets of data, consistently generating long measured sequences poses sufficient barriers to the proliferation of undisclosed synthetic data.

## Introduction

Applications of artificial Intelligence (AI) are becoming increasingly integrated into modern scientific research. Recent studies suggest that up to 84% of researchers are using AI in their professional workflows (*1*). AI is used for summarizing scientific information from published literature and for automating routine tasks, such as coding, data retrieval and analysis (*2*). This widespread adoption is heralded by major advances in state-of-the-art AI and its applications that range from breakthroughs in protein structure prediction (*3*) to the emergence of large-scale foundation models (*4*, *5*). In some cases, AI is credited for assistance in generating scientific hypotheses and experimental protocols, as well as for enabling semi-autonomous laboratories (*6*, *7*). Progress is being made in AI-driven theoretical physics calculations (*8*) and mathematical proofs (*9*).

Among potentially beneficial applications is the use of AI for synthetic data that mimics real-world data but is not produced by experiment or survey (*10*). Synthetic data can be helpful, provided it accurately reproduces the target statistical properties and relationships, in cases where data is scarce, sensitive or too difficult to collect. For example, in medical research, the original dataset can be rebalanced with synthetic data to account for underrepresented groups or medical conditions; to test for and mitigate bias in health risk prediction tools; or to address privacy in applications that rely on face recognition (*11*, *12*).

Successful use cases are contrasted with various limitations of AI technology leading to sustained skepticism (*13*). Scientists are growing concerned that AI can be used in questionable research practices (*14*, *15*). Issues typically arise when large language model (LLM) products are misrepresented as human products, as is the case with AI-generated referee reports (*16*). Unreliable AI products create challenges, such as when AI suggests to cite non-existent papers (*17*), or when AI-generated images contain inaccuracies known as “AI hallucinations” (*18*, *19*). An emerging threat is from AI text in scientific publications or abstracts (*20*, *21*), which can dilute the credibility of the scientific record. These concerns also apply to synthetic data generated by AI, but described as results of actual experiments, studies or numerical simulations (*22*).

Even in the fields perceived as data-heavy, such as physics, actual volumes of data included with a publication are relatively small and thus easily fungible by AI, while their verification is often limited to an eye inspection of the figures (*23*). While data fabrication and falsification predate the mass adoption of AI, prompt engineering simplifies this dubious task. With time and continued training, the ability of AI to mimic data is likely to grow making it urgent to address this challenge now.

In this paper, we address the following questions: Can consumer AI synthesize data capable of mimicking hallmark quantum physics experiments? How plausible can the resulting data be compared to real measurements? And, most importantly, how can AI fabrications be detected? We use the Data Analyst app within OpenAI’s ChatGPT. We focus on superconducting transmon qubits (*24*), a leading hardware platform in quantum computing, on topological nanowires which can host Majorana Zero Modes (*25*) relevant for topological quantum computing (*26*), and on quantization effects found in semiconductor quantum dots, quantum wells and quantum point contacts (*27–29*).

We demonstrate that AI is capable of creating new tabulated datasets as well as augmenting existing data to alter the interpretation towards a more desirable hypothesis. Starting with idealized calculations from basic physics models and formulas, we add realistic experimental artefacts such as high frequency noise, or sudden jumps in the feature position within the data map. We are able to endow the noise with the spectrum taken from a common laboratory lock-in amplifier. On the other hand, we observe and document how ChatGPT struggles to maintain internal consistency that is expected when conducting longer series of measurements on the same sample. Based on this, we argue that increased sharing of the original data and other primary research materials can provide powerful means of data authentication in the era of AI.

**Methods**

We use the Data Analyst application within OpenAI's ChatGPT environment (*30*). The version at the time of writing is based on ChaptGPT 5.4, some of the examples shown in the supplementary materials used older models, namely 4o and 5.2. Data Analyst is an application designed to work with uploaded data. It has access to a large array of Python libraries for data analysis, including Pandas and Matplotlib for visualization. Upon receiving a prompt, the application generates and executes a Python script applied to the uploaded data. Data Analyst can parse tabulated data in some of the common formats (such as xls, csv), and can also handle lab-specific formats if the file structure is specified through prompts or if uploaded data files contain metadata. The application is able to apply common mathematical transformations, perform statistical analysis and visualize the results.

In addition to the analytical functions, "file generation and manipulation" is also listed among the applications (*30*). Thus Data Analyst shows capacity to interpret a prompt expressed in specialized scientific terms, such as "plot Coulomb blockade data for a quantum dot". Prompted about a specific physics topic, the model may return a theoretical description accompanied by a script which is attempting to simulate a physics system in the prompt. For example, the code may contain differential equations (see supplementary information), and numerically solve them in a sandbox environment. We confirmed this by independently re-running the Python code that Data Analyst returns. Synthetic data can then be further manipulated via prompts, for instance to make it appear more realistic through adding noise, signal jumps and other non-idealities. Or, on the contrary, to make the synthetic data fit better to a known physics model or hypothesis. Synthetic data can be downloaded in a chosen format, or as a graphic. Data Analyst can output data in a desired format. For instance it can overwrite a column into the original file from an actual measurement.

**Results**

1. Fully synthesized data for a superconducting transmon qubit

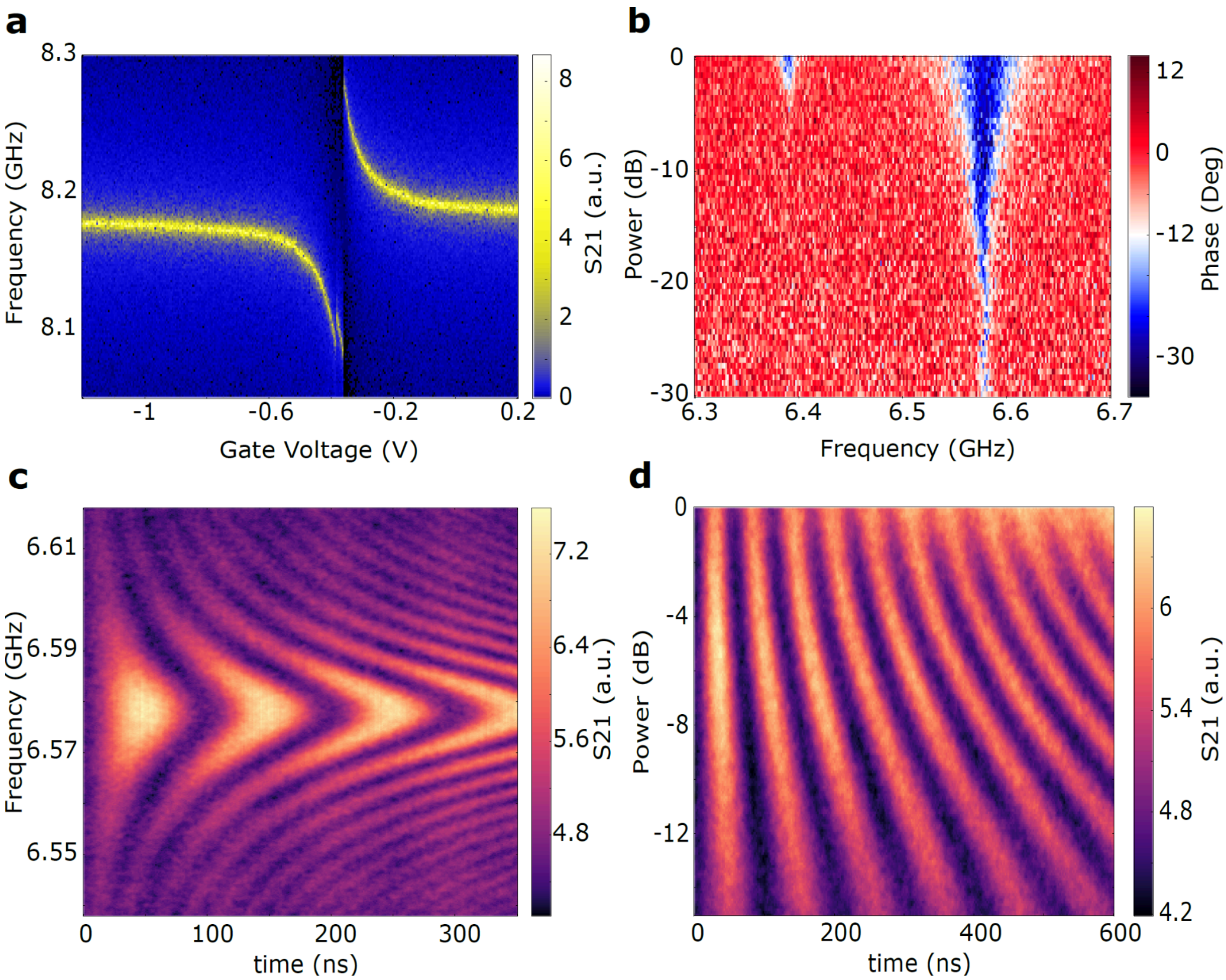

**Figure 1**: *Synthetic transmon data. a) Gate voltage dependence of single tone response of a superconductor-semiconductor transmon coupling to a λ/4 coplanar waveguide resonator with frequency 8.182 GHz, coupling quality factor 2000. b) Two-tone spectroscopy as a function of power showing simulated* $f_{01}$ *=6.578 GHz and* $f_{02}/2$ *resonances using qubit anharmonicity -380 MHz. c) Rabi oscillations "chevron" pattern using relaxation time* $T_1$ *= 5.6 µS. d) Rabi oscillations as a function of power.*

In Figure 1 we demonstrate a set of synthetic data for a superconducting transmon qubit (*31*), an electronic device that is the basis for several prominent quantum computing projects (*32*, *33*). Improvement of transmon qubit metrics such as coherence times remains a major goal in quantum computing (*34*), as it directly translates into the higher performance of prototype quantum processors. Therefore, papers reporting basic characterization of transmons frequently appear in the literature. The panels of the figure are consistent with data obtained on a single realistic transmon device. They represent a set that is typically shown in transmon characterization papers, and answer a lot of the questions that a referee or a reader may have about the device.

Transmon consists of a Josephson junction shunted by a capacitor. At low temperatures, typically below 1 degree Kelvin, this circuit demonstrates properties of the quantization of energy levels and superpositions of charges between capacitor plates. A standard transmon is a well-understood object that can be described theoretically (*31*). Having written down the transmon Hamiltonian, it is possible to calculate the energy spectrum, and the time evolution of the transmon quantum states by solving the Schrodinger equation. Such calculations done using ChatGPT return results in theoretical model parameters such as state occupation probabilities (not shown here).

We prompt Data Analyst to synthesize from scratch the data that could serve as a fair representation of laboratory measurements on a gate-voltage tunable transmon qubit coupled to a coplanar waveguide microwave resonator, similar to Ref. (*35*). ChatGPT is asked to express data in the form expected for dispersive readout, such as the amplitude S21 and the phase of the transmitted microwave signal. We separately prompt the model to add non-ideal behavior, such as noise and charge jumps.

All panels of Figure 1 may have originated from the same transmon sample: they convey a consistent set of device parameters. In Fig. 1a, we show the single microwave tone spectroscopy which demonstrates a level anticrossing due to the interaction between a gate-voltage tunable transmon and a microwave resonator. Charge jumps lead to abrupt shifts in the spectral line best visible immediately to the left of the anticrossing (Gate voltage of -0.4 V). Fig. 1b is focused on the two-tone spectroscopy with transmission phase as a signal. It displays the primary ground-to-excited state qubit transition (right dip) which extends to lower power, and the higher order two-photon transition from the ground to the second excited state (left dip), which requires higher power to become observable. The jitter in the peak position is an added feature of this synthetic dataset. Figs. 1c and 1d show two different representations of the iconic driven quantum (Rabi) oscillations with added measurement noise.

In the supplementary materials we provide examples of synthetic data for several other commonly studied systems such as quantum dots and quantum point contacts. Similar to the transmon qubit, the basic mathematical models describing them are well-established, and this makes it possible for LLM to translate a prompt into a Python code for data synthesis. It is straightforward to mimic individual line traces, such as an exponential decay or a sinusoidal oscillation which describe a vast array of physics phenomena. Many quantum computing papers report results in the form of a density matrix which are sets of numbers constrained by relations such as normalization. We illustrate ChatGPT's ability to synthesize such data using the example of an entangled three-qubit Greenberger–Horne–Zeilinger **(**GHZ) state (*36*) (see supplementary materials). We note that some aspects of the physics of real devices, such as transitions to higher transmon energy levels, may not be accounted for by ChatGPT, offering a physics-based pathway for detecting synthetic data.

2. Augmented experimental data - the Majorana peak

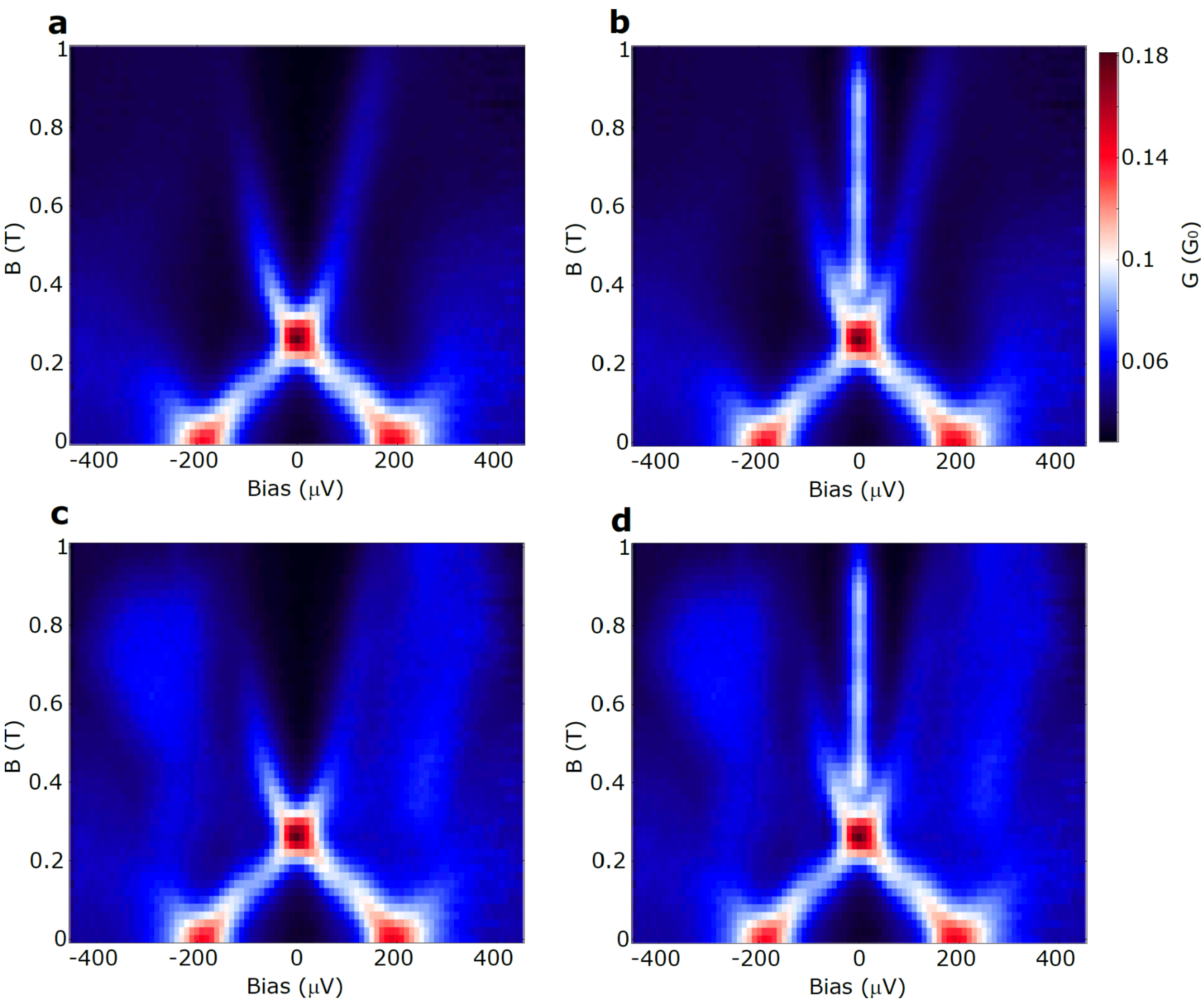

**Figure 2**: *AI-augmented Majorana nanowire data. a) Real experimental tunneling spectroscopy data from Ref. (37). b) Panel a) augmented with a zero-bias peak. c) Panel a) augmented with higher finite-bias conductance. d) Augmentations from panels b) and c) combined.*

In Figure 2, we demonstrate how experimental data can be augmented using AI. For this example we start with real tunneling spectroscopy data from Ref. (*37*) on the so-called Majorana nanowires. In these devices, superconductor nanowires are coated with a layer of a superconductor. This induces quantized states in the nanowire which can be detected by resonant tunneling across a potential barrier. The features manifest as peaks on a smoother background. Fig. 2a demonstrates the non-augmented data with two of the states originating at a source-drain bias voltage of +/- 200 μV, shifting to lower bias with magnetic field B and crossing at zero bias around B~0.3 T (red blob in the center). While these are interesting data, they are described as “trivial” because the resonances are sensitive to magnetic fields.

We prompt the Data Analyst to add to the real experimental data a magnetic-field insensitive peak at zero bias voltage (Fig. 2b). This may correspond to the presence of Majorana Zero Modes (*25*), the non-Abelian anyons that can facilitate topological quantum computing. We use a series of prompts to make sure the peak blends smoothly with the original data. The augmentation only alters a narrow vertical strip several pixels wide in the center of the figure.

By requesting that the peak starts near the crossing point and extends to higher magnetic fields, we invite the discussion of whether the data correspond to a phase transition into a topologically superconducting state, a new state of matter (*38*). A phase transition is theoretically expected to be accompanied by a closing and a re-opening of a superconducting gap (*39*). To emulate this effect, we prompt AI to augment the original experimental data with larger signals outside of the central crossing of the two lines (Fig. 2c vs. Fig. 2a). The tunneling conductance is smoothly and subtly enhanced at biases larger than those of the crossing resonances at each magnetic field. Combining the augmented zero-bias peak with the closing gap augmentation, we produce data in Fig. 2d, which is a synthesis of Figs. 2b and 2c. A trivial data set is transformed into a one suggestive of a major physics discovery.

While Figure 2 shows a dramatic example, it also demonstrates that more subtle augmentation is possible for beautification purposes or to nudge the experimental data closer to the desired look. This raises a concerning possibility of a workflow in which every data file is augmented by AI in real time shortly after data acquisition. If only a small fraction of pixels in a data array are modified, as is the case in Fig. 2b, it may be challenging for any AI detection algorithm to identify that augmentation has been performed. However, the strategy would have its limits. As the quantum device is further interrogated and tuned, larger and larger augmentations will be required to keep the new data consistent with the preferred look or theory. In mesoscopic devices each of which has a “fingerprint” of features in the data (e.g. peaks, dips, shifts, oscillations), this would be difficult to implement without creating an increasing number of inconsistencies.

3. Mimicking a common lock-in amplifier

For a lock-in, the signal can be split into several components such as the X and Y quadratures. Readings may be transmitted and recorded with a particular number of digits and with a certain acquisition time. All these characteristics need to be consistent in the synthetic data, and any discrepancies may be detected through the presence of computational artefacts, rounding, and differences in noise in different columns.  Some data acquisition scripts also record instrument’s settings, such as the averaging time, the bandwidth of any filters. However, files that satisfy these types of constraints can be synthesized by providing AI with an example of a real data file.

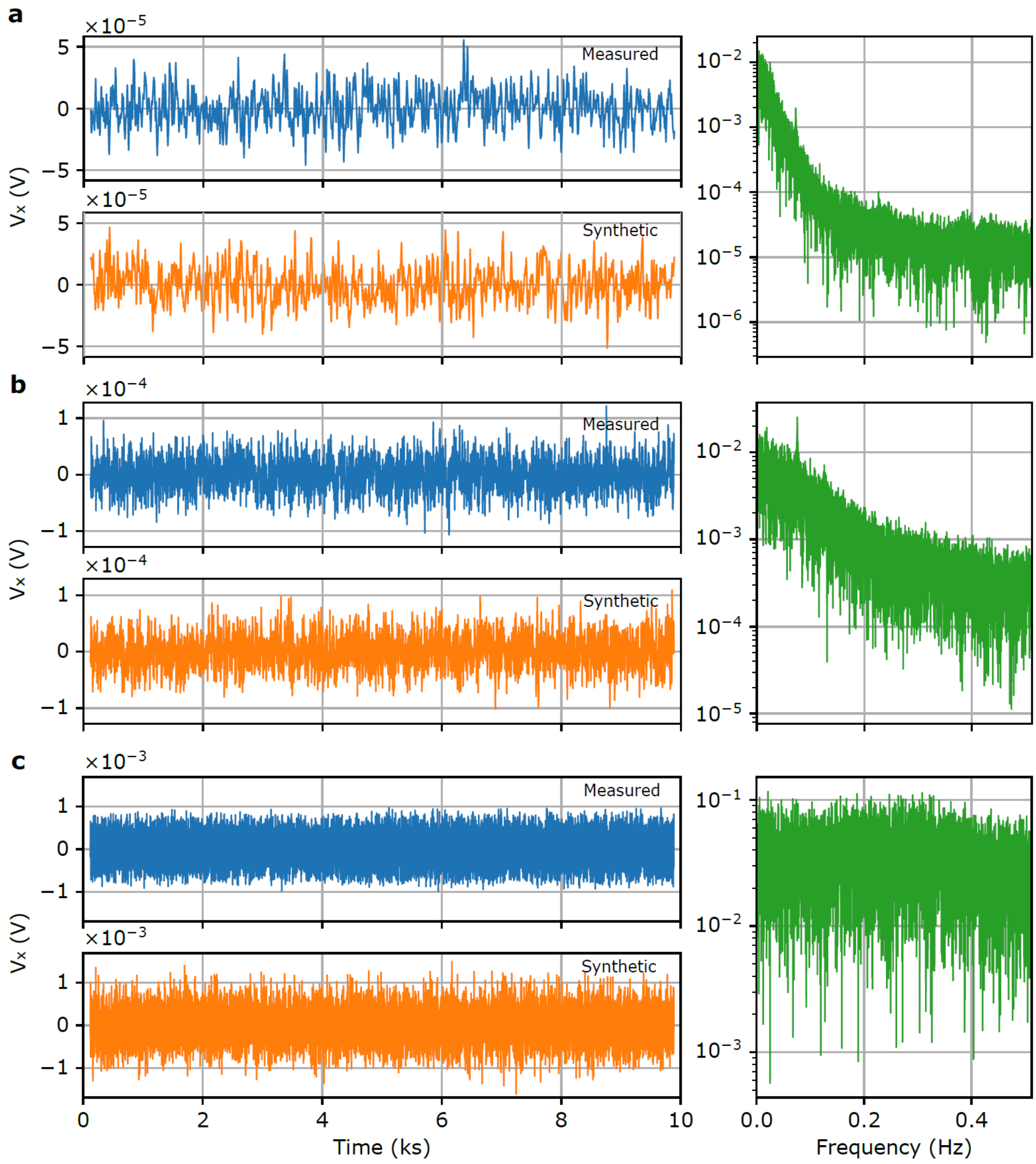

**Figure 3:** *Real vs. synthetic idle noise of a lock-in amplifier. Measured, fast fourier transformed and synthetic data for x-component of voltage Vx from a Stanford Research Systems SR830 lock-in amplifier using the following settings: a) integration time constant 300 ms, filter 24 dB/octave b) 100 ms, 18 dB/octave c) 100 ms, 6 dB/octave. Data plotted in virtual time (1s between data points).*

We load into Data Analyst several time traces of idle measurements, recorded off a common Stanford Research Systems lock-in amplifier model SR830 found in many physics labs. Each trace is collected for 10-15 minutes at different lock-in settings. For ease of comparison between Figs. 3a, 3b and 3c, we plot data as a function of virtual time, which runs at 1s between data points (real time is recorded in the raw data files available on Zenodo). We then ask the model to perform a fast Fourier transform (FFT) of the signal vs virtual time. Subsequently, we prompt the model to generate a new signal that would result in the same FFT, by randomizing the wave amplitudes and phases. The FTT of the synthetic signal is identical to the FFT of the original signal and therefore is not displayed.

Noise signals such as those shown in FIgure 3 can be directly added to relevant synthetic data for a quantum device. Alternatively, the FFT data can be used to define the noise model. This can be used to improve upon the fully synthetic data such as those in Figure 1 where ChatGPT came up with its own noise model not informed by typical noise from any common laboratory equipment.

**Discussion and Conclusions**

We demonstrate that at the time of writing consumer-level generative AI systems are capable of synthesizing the types of signals and patterns that appear in a significant number of experimental papers on quantum devices. LLMs such as ChatGPT have already been trained with enough information on the methodologies underlying these papers. AI does not necessarily need to learn from the large body of domain-specific data to generate realistic examples in this field because it can calculate them from a basic set of physics equations. Indeed, physicists often select data that can be fit by some of the simplest models. A plausible dataset can be synthesized starting with an intuitive prompt such as “generate data for a qubit”, because a qubit is a textbook two-level quantum system. A quantum dot is an undergraduate physics particle-in-a-box problem at its core.

AI applications such as Data Analyst can access the tools for signal processing. The generated output can be iteratively refined using more data-centric instructions, such as prompting the tool to introduce realistic noise, line broadening, instabilities in device parameters, to suppress selected signal frequencies, or modify more detailed signal characteristics to match experimental behaviour more closely. What makes research in quantum devices fungible by AI is that only a small fraction of results are typically reported in a publication where 5-10 datasets are selected for figures out of hundreds or even thousands. AI is capable of creating paper’s endpoint figures that are visually plausible and consistent with each other.

These capabilities substantially reduce the expertise and effort previously required to synthesize technically convincing research data, thus creating a concern that fabricated dataset will be increasingly more difficult to distinguish from authentic experimental measurements. It has always been possible to synthesize data without AI, by creating a code or “by hand”, i.e. by drawing a graph or writing down numbers. However the ease with which generative AI produces realistic outputs is unmatched. A conversational interface can combine domain knowledge, signal generation, numerical manipulation, and iterative refinement in a single workflow.

For the time being, a powerful solution to this problem is to increase the sharing of the original data, code, sample preparation and other experimental methods (*40*). Examination of the full suite of primary research materials likely provides enough evidence that it was not AI generated or fudged in another. An experimental study may last months and combine automatic data acquisition with human inputs to direct and modify the experimental strategy, conducted over days, weeks and months. The results are stored in thousands of files, adding up to between hundreds of megabytes and terabytes. For each file, multiple columns are recorded, accompanied by metadata with the measurement setup information, readings of temperatures, magnetic fields, timestamps, control parameters and their units, and amplification factors.

It is presently beyond the Data Analyst's capabilities to synthesize such data in a consistent fashion. First, the context window cannot accommodate gigabyte-scale datasets, and second, the model does not maintain consistency over multiple iterations, meaning that the model loses track of all the manipulations done on a dataset. In our testing, the model wasn't capable of rolling back to the previous version of the data consistently and reliably, if a prompt resulted in an undesired modification. A branching feature exists that allows the chat to be resumed from an earlier point. However, this creates a proliferating tree of chat branches that becomes increasingly difficult to keep track of.

We considered other approaches to verifying data provenance and found them lacking compared with data sharing, for reasons we outline below. Statistical methods for identifying anomalous or potentially fabricated data have been developed in statistics-heavy fields such as psychology, clinical trials and biostatistics prior to the advance of AI (*41–44*). These approaches may be helpful in detecting certain types of issues, such as unusual formatting or implausible distribution of residues. However, this may not be sufficient for detecting AI-assisted fabrication (*45*) which can take these aspects into account.

The development of AI detection tools for synthetic data is likely to result in a race between new synthesis and detection approaches, which has arguably begun in other domains such as synthetic text and image detection (*46*). Our results demonstrate that augmentation of files by altering a small number of key data points, without modifying the bulk of the dataset, can be an effective and difficult-to-detect way of supporting a preferred hypothesis (Figure 2). Adding realistic noise that corresponds to real laboratory instruments (Figure 3) would make AI-detection even harder.

LLM vendors could implement watermarks that are added to the synthesized data, and this could help increase the barrier of using undisclosed synthesized data, though workarounds are likely to be found through prompt engineering, switching to a different vendor or scrambling the data additionally post-synthesis. Encoding data in a blockchain could help verify the absence of post-processing, but real-time augmentation can still take place.

While data sharing is typically associated with open science which also advocates for open source formats, we acknowledge that in the context of data provenance, proprietary data formats may offer means of verification. Since such formats are less transparent, current AI tools are unlikely to generate internally consistent files that would pass format validation, let alone reproduce the specific noise characteristics and metadata of a real instrument.

The model performance has improved considerably between 2023 when our initial tests were done and 2026, when this manuscript was written. It is expected that in the future prompt interpretation and the flexibility of operating with datasets will further improve. Targeted training on data found in the literature may further enhance the similarity between the synthetic and real data in the future. However, at the present moment, our observations illustrate that sharing larger volumes of data is an effective means against undisclosed synthetic, as well as fabricated or manipulated data.

In the longer term, ethics training will likely incorporate ethical uses of AI across various academic and scholarly applications, and this should include synthetic data and data processing. We recommend that this training not be viewed separately but is treated as part of the broader education in how to be a good scientist. The topics of data verification are closely related to the depth of knowledge in a specific field, the ability of a scholar to evaluate the claims of a paper and estimate the likelihood of their replicability, including whether or not a sufficient volume of data has been shared to verify the conclusions of a study. Ultimately, independent replication remains the strongest method to prove the validity of a scientific discovery. However, the hard work of replication and experimental follow-up should not be spent on undisclosed synthetic data-based claims.

Data Availability Statement. Data of the main text figures, plots generated by ChatGPT and screenshots of some of the chats with Data Analyst are available on Zenodo https://doi.org/10.5281/zenodo.20514192.

Acknowledgements: We thank M. Gomanko and S. Mudi for assistance in collecting lockin noise data. Work on synthesizing data with AI supported by the Department of Energy Office of Science grant DE-SC0022073.

AI use disclosure: ChatGPT DataAnalyst based on several versions of the model (2023 to 2026) was used to synthesize data used in figures of this paper. No large language models were used to write or edit the text.

**Supplementary materials**

**Realistic AI-synthesized quantum device data is identifiable through full data sharing**

S. M. Frolov [1] and O. V. Kravchenko [2,3]

**Additional Example 1: Quantum Point Contact**

Initial prompt that generated the single conductance trace below (scroll-through chat with Data Analyst on Zenodo):

> Can you generate a synthetic dataset for a conductance trace of a quantum point contact, including minor quantum interferences on top of conductance quantized plateaus due to backscattering into the qpc? And visualize the trace

Data Analyst: I'll generate a physically plausible synthetic QPC trace with smooth conductance steps plus small interference fringes/backscattering features, then plot it and save the dataset.

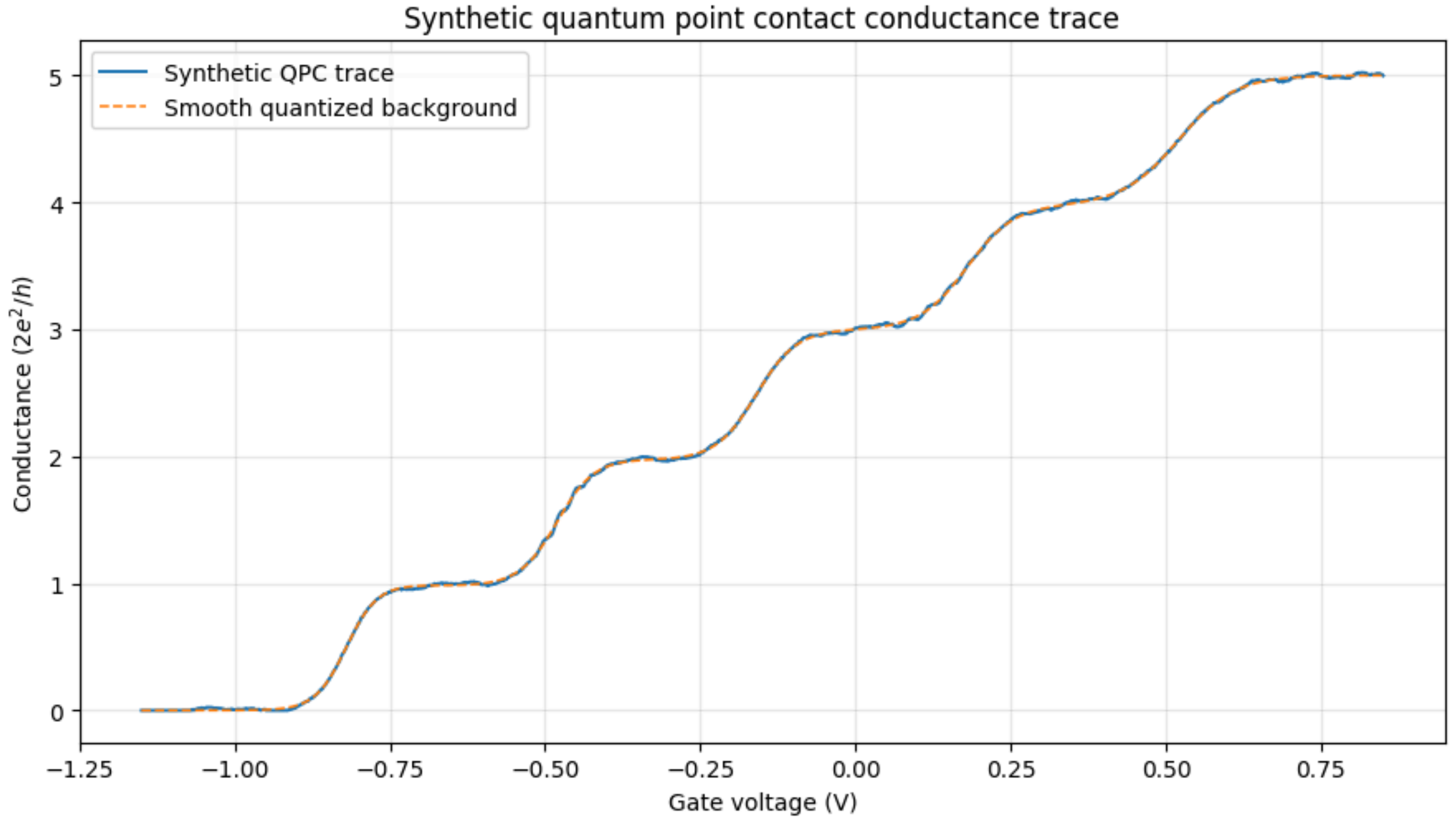

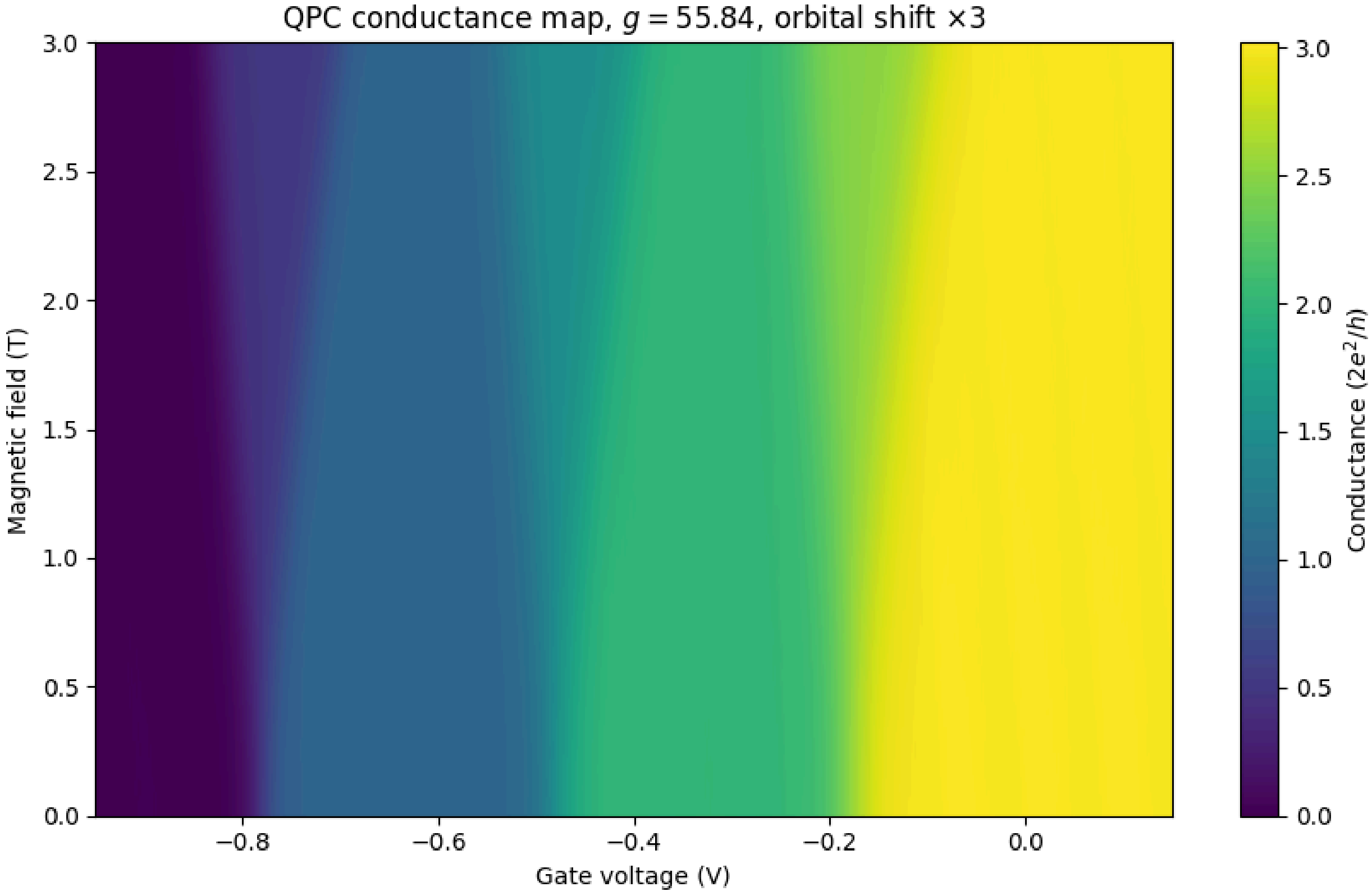

QPC conductance map, $g = 55.84$, orbital shift ×3
Magnetic field (T)
Gate voltage (V)
Conductance ($2e^2/h$)

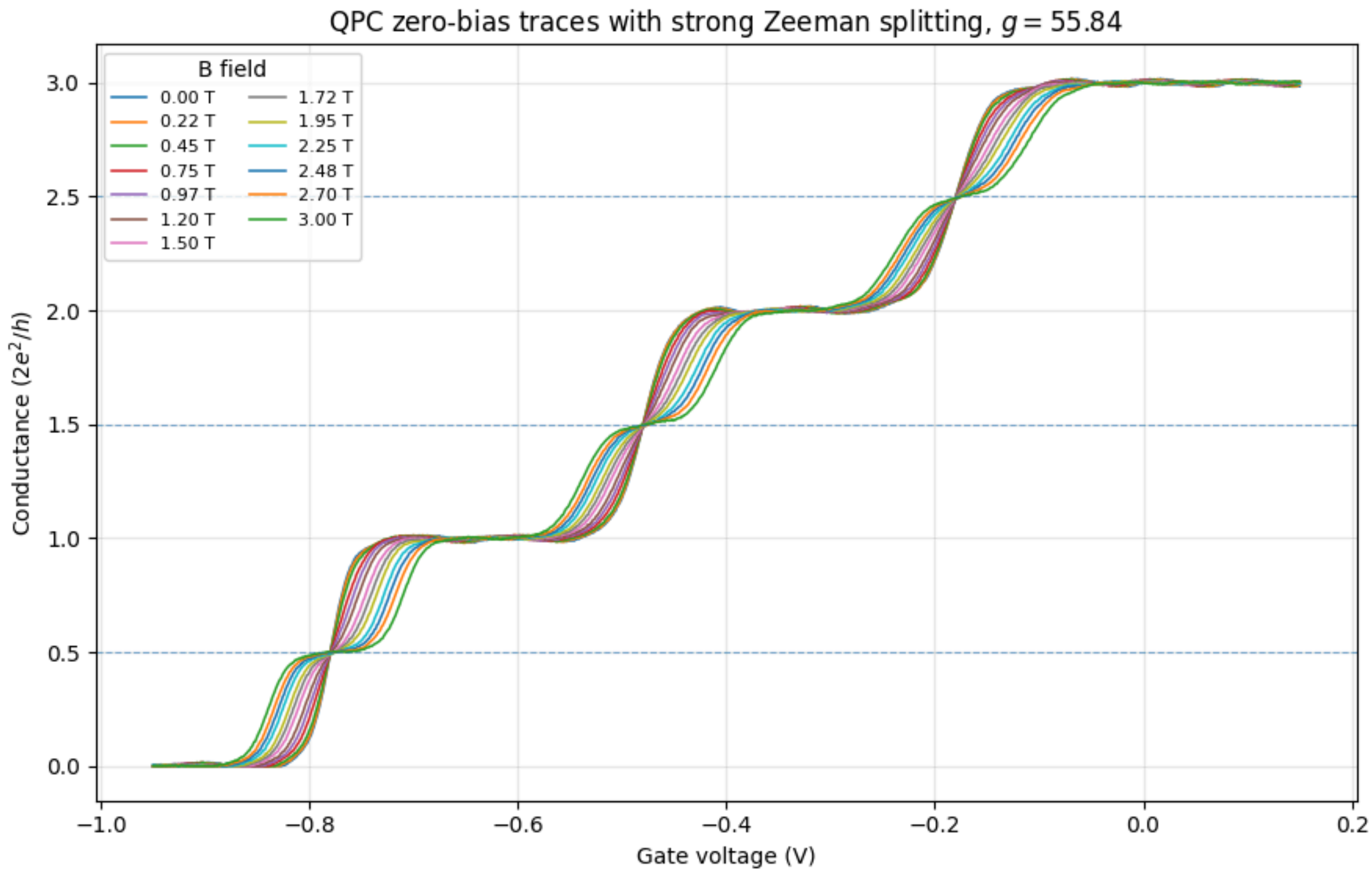

QPC zero-bias traces with strong Zeeman splitting, $g = 55.84$
B field
0.00 T
0.22 T
0.45 T
0.75 T
0.97 T
1.20 T
1.50 T
1.72 T
1.95 T
2.25 T
2.48 T
2.70 T
3.00 T
Conductance ($2e^2/h$)
Gate voltage (V)

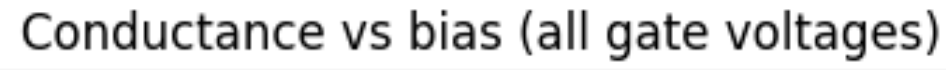

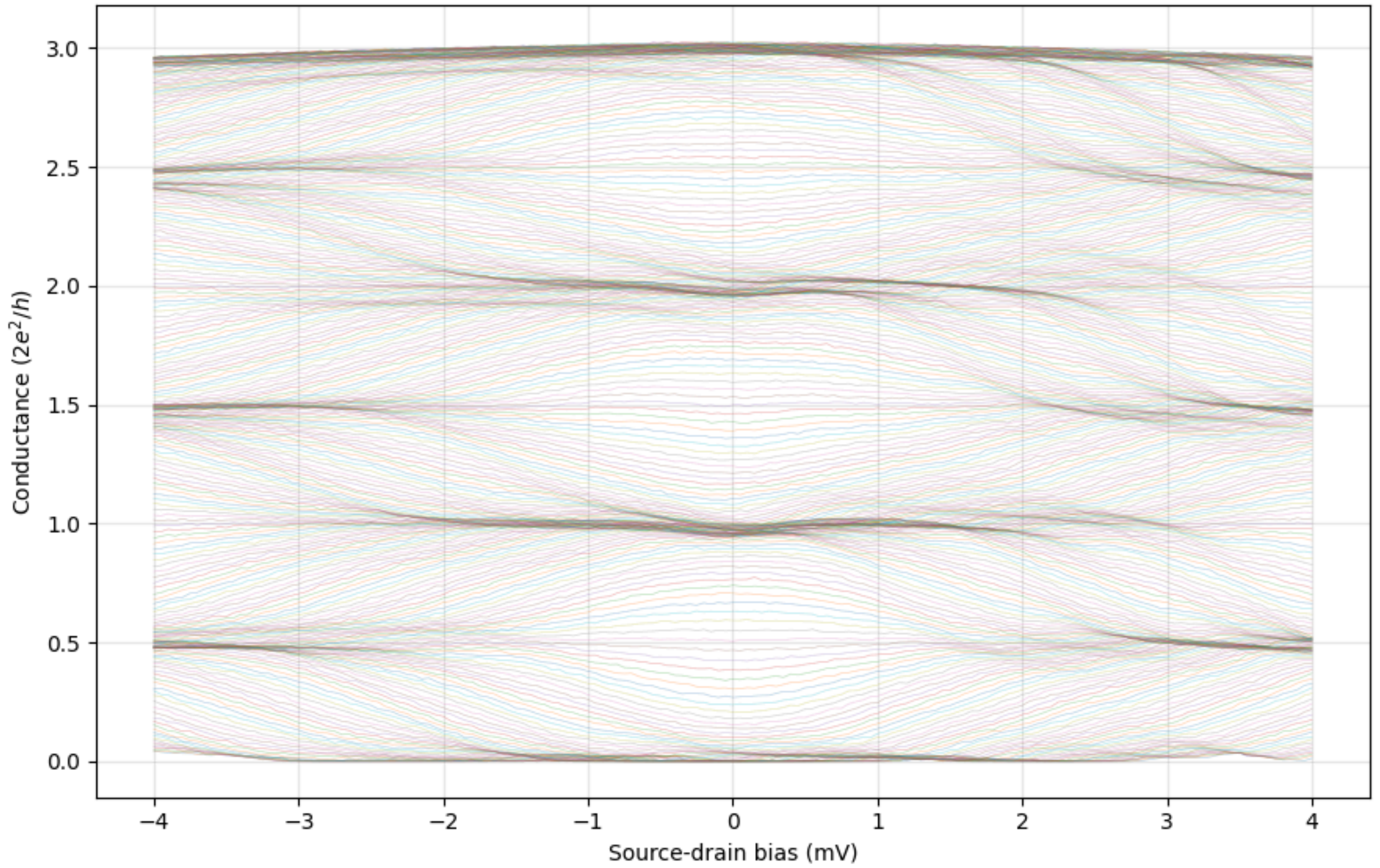

**ChatGPT-generated code for the Quantum Point Contact Conductance vs bias “waterfall” plot for different gate voltage settings:**

```
import numpy as np
import matplotlib.pyplot as plt

rng = np.random.default_rng(11)

# Axes
Vg = np.linspace(-0.95, 0.15, 360)     # gate voltage, V
Vsd = np.linspace(4.0, -4.0, 161)      # source-drain bias, mV

# First three QPC modes
thresholds = np.array([-0.78, -0.48, -0.18])
widths = np.array([0.018, 0.022, 0.026])
alpha = 0.060   # gate-voltage equivalent shift per mV bias

G2d = np.zeros((len(Vsd), len(Vg)))

for i, vb in enumerate(Vsd):
    bias_shift = alpha * abs(vb) / 2
    G_line = np.zeros_like(Vg)

    # Finite-bias conductance from averaged source/drain transmissions
```

```
    for vt, w in zip(thresholds, widths):
        T_plus = 1 / (1 + np.exp(-(Vg - (vt - bias_shift)) / w))
        T_minus = 1 / (1 + np.exp(-(Vg - (vt + bias_shift)) / w))
        G_line += 0.5 * (T_plus + T_minus)

    # Weak zero-bias/backscattering correction
    zba = -0.035 * np.exp(-(vb / 0.45) ** 2) * (
        np.exp(-((Vg + 0.62) / 0.16) ** 2)
        + 0.8 * np.exp(-((Vg + 0.32) / 0.16) ** 2)
    )

    # Small quantum interference fringes
    plateau_weight = np.exp(-(np.abs(np.gradient(G_line, Vg)) / 7.0) ** 2)
    phase = 2*np.pi*(10.5*Vg + 0.42*vb + 1.6*Vg**2 + 0.035*vb**2)

    fringes = plateau_weight * (
        0.018*np.sin(phase)
        + 0.007*np.sin(2.4*phase + 0.6)
    )

    # Small conductance suppression at high bias
    high_bias_suppression = -0.018 * (abs(vb) / max(abs(Vsd)))**1.5 * G_line

    # Experimental-like noise
    noise = rng.normal(0, 0.0025, size=Vg.size)

    G2d[i, :] = np.clip(
        G_line + zba + fringes + high_bias_suppression + noise,
        0,
        None
    )

# Plot conductance vs bias for all gate voltages
plt.figure(figsize=(9, 6))

for idx in range(len(Vg)):
    plt.plot(Vsd, G2d[:, idx], linewidth=0.4, alpha=0.35)

plt.xlabel("Source-drain bias (mV)")
plt.ylabel(r"Conductance ($2e^2/h$)")
plt.title("Conductance vs bias (all gate voltages)")
plt.grid(True, alpha=0.3)
plt.tight_layout()
plt.show()
```

**ChatGPT-generated explanation of the code:**

The model builds the conductance in units of 2e^2/h. Each plateau comes from a smooth logistic transmission step, representing the opening of one QPC subband. At finite source-drain bias, each subband edge is split into two effective thresholds, one for each lead chemical potential, and the conductance is averaged between them.

Then I added small corrections to make it look more experimental: weak backscattering near zero bias, gate- and bias-dependent interference fringes, mild high-bias suppression, and small random measurement noise. The final plot simply takes vertical slices at every gate voltage and plots conductance versus source-drain bias with no vertical offset.

**Additional Example 2: Quantized zero-bias conductance peak inside a hard induced gap due to “Majorana fermions”**

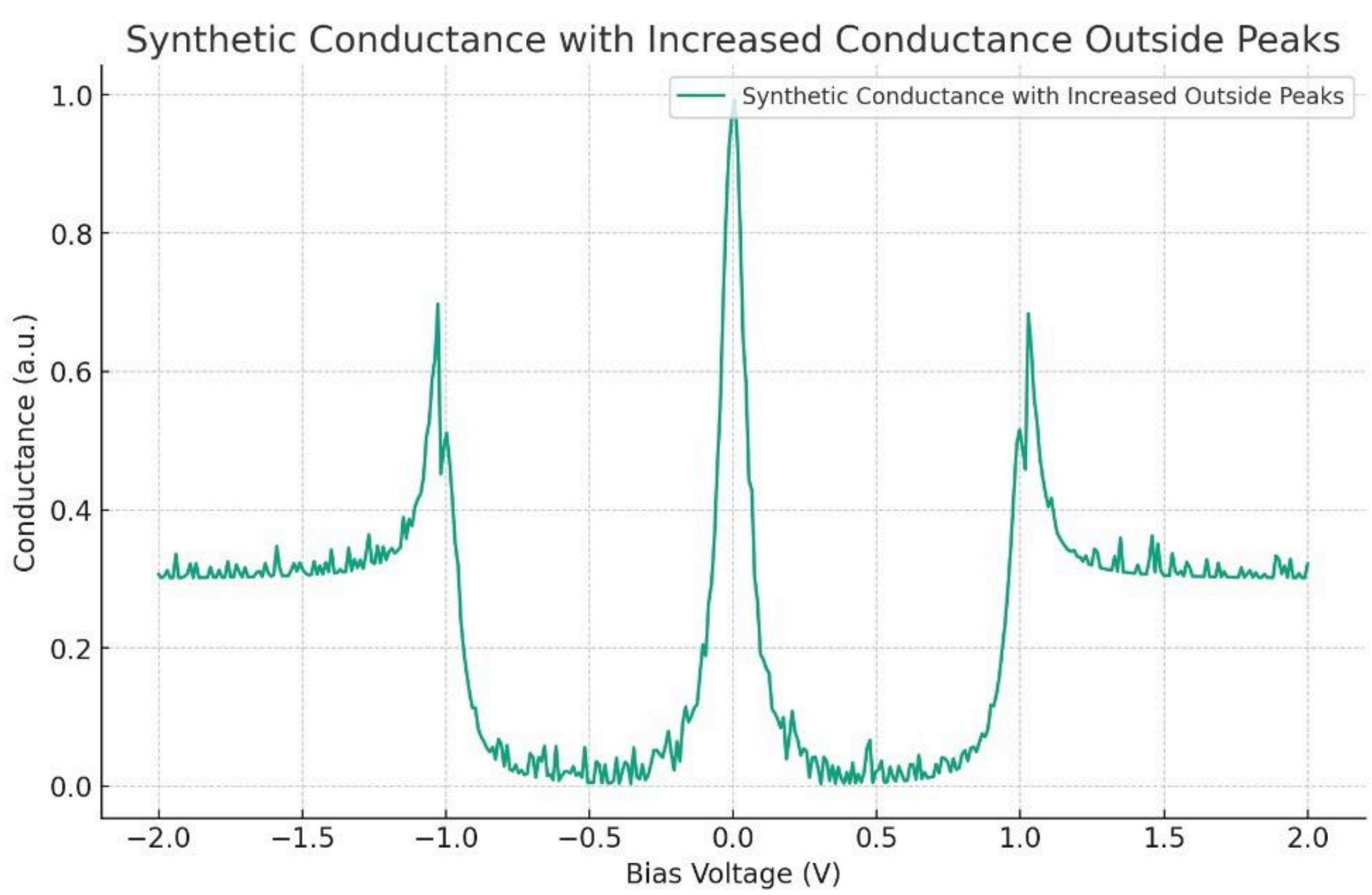

**Additional Example 3: Quantum Dot Coulomb Diamonds with asymmetric coupling to source and drain contacts, variations in orbital energies and with added charge jumps and noise**

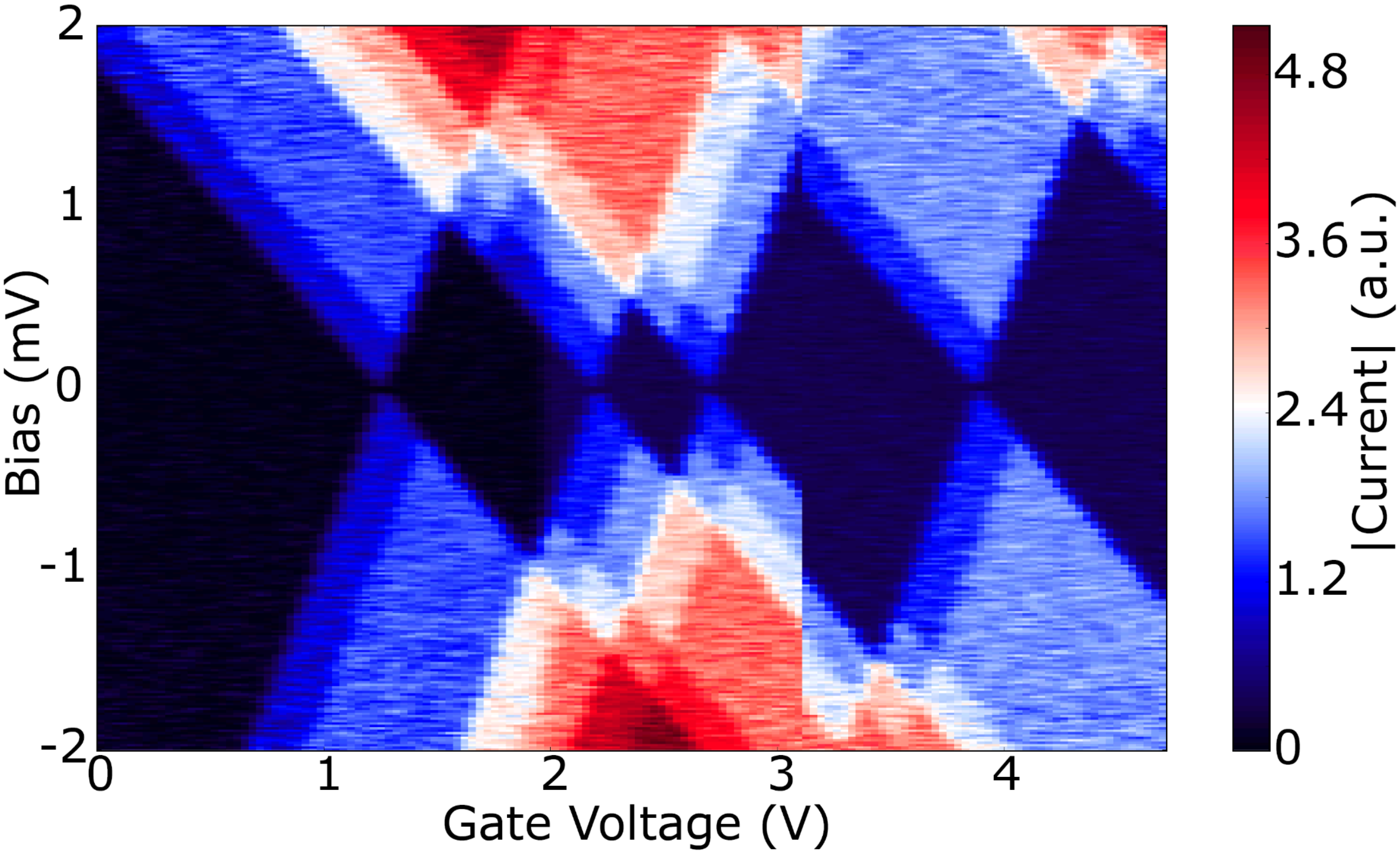

**Additional example 4: Greenberger–Horne–Zeilinger (GHZ) three qubit state tomography**

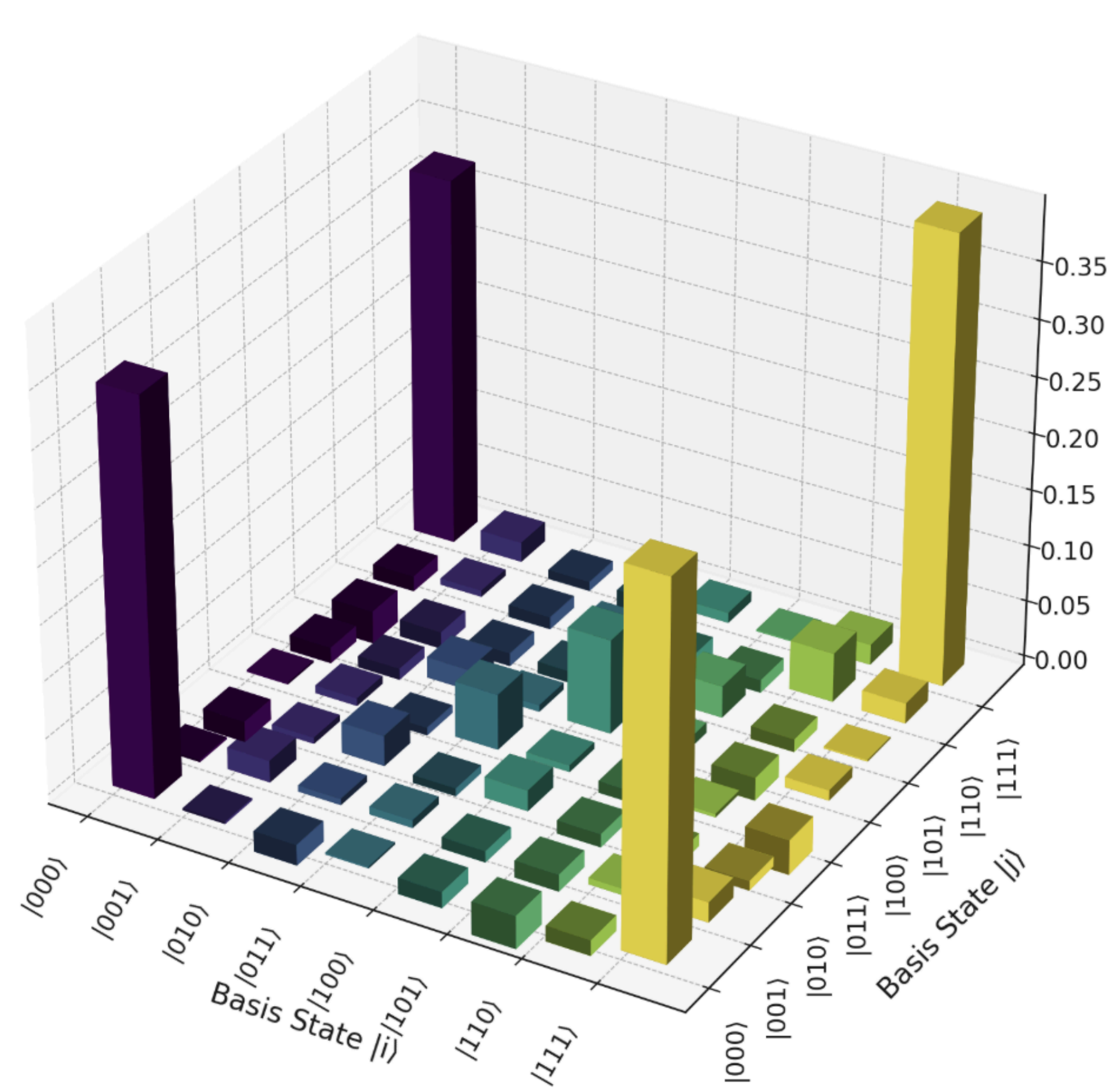